\documentclass[aoas,MSNbibl,nameyear,seceqn,rotating,dvips]{arximspdf}
\usepackage{dcolumn}

\doi{10.1214/11-AOAS526}
\volume{6}
\issue{2}
\pubyear{2012}
\firstpage{497}
\lastpage{520}

\makeatletter
\renewcommand{\epsilon}{\varepsilon}
\newcolumntype{e}[1]{D{.}{}{#1}}
\newcommand{\n}[1] {\mathbf{#1}}
\makeatother

\begin{document}
\begin{frontmatter}

\title{A Bayesian model averaging approach for observational
  gene expression studies}
\runtitle{A BMA approach for observational micorarray studies}

\begin{aug}
\author[A]{\fnms{Xi Kathy} \snm{Zhou}\corref{}\thanksref{t1}\ead[label=e1]{kaz2004@med.cornell.edu}},
\author[B]{\fnms{Fei} \snm{Liu}\thanksref{t2}\ead[label=e2]{feiliu@us.ibm.com}}
\and
\author[C]{\fnms{Andrew J.} \snm{Dannenberg}\thanksref{t3}\ead[label=e3]{ajdannen@med.cornell.edu}}
\runauthor{X. K. Zhou, F. Liu and A. J. Dannenberg}
\affiliation{Weill Medical College of Cornell
  University, IBM Watson Research Center and Weill Medical College of Cornell
  University}
\address[A]{X. K. Zhou \\
  Division of Biostatistics and Epidemiology \\
  Department of Public Health \\
  Weill Cornell Cancer center \\
  Weill Medical College of Cornell University\\
  New York, New York 10065\\USA\\
  \printead{e1}} %adresu isvedimo komanda gale!
\address[B]{F. Liu \\
  Department of Business Analytics\\\quad and Mathematical Sciences\\
  IBM Watson Research Center\\
  Yorktown Heights, New York 10598\\USA\\
  \printead{e2}}
\address[C]{A. J. Dannenberg \\
  Department of Medicine \\
  Weill Cornell Cancer center \\
  Weill Medical College of Cornell University\\
  New York, New York 10065\\USA\\
  \printead{e3}}
\end{aug}
\thankstext{t1}{Supported in part by a pilot award from
  the Clinical and Translation Science Center at Weill Cornell Medical
  College through the National Institute of Health (UL1-RR024996) and
  National Cancer Institute 1R21CA133260-01A2.}
\thankstext{t2}{Supported in part by the University of Missouri
  Research Board Award.}
\thankstext{t3}{Supported in part by funding from the Flight
  Attendants Medical Research Institute.}

% HISTORY:
\received{\smonth{6} \syear{2011}}
\revised{\smonth{10} \syear{2011}}

% ABSTRACT
\begin{abstract}
Identifying differentially expressed (DE) genes associated with
a~sample characteristic is the primary objective of many microarray
studies. As more and more studies are carried out with observational
rather than well controlled experimental samples, it becomes
important to evaluate and properly control the impact of sample
heterogeneity on DE gene finding. Typical methods for identifying DE genes
require ranking all the genes according to a preselected
statistic based on a single model for two or more group
comparisons, with or without adjustment for other covariates. Such
single model approaches unavoidably result in model
misspecification, which can lead to increased error due to bias
for some genes and reduced efficiency for the others. We evaluated
the impact of model misspecification from such approaches on
detecting DE genes and identified parameters that affect the
magnitude of impact. To properly control for sample
heterogeneity and to provide a flexible and coherent framework for
identifying simultaneously DE genes associated with a single or
multiple sample characteristics and/or their interactions, we
proposed a~Bayesian model averaging approach which
corrects the model misspecification by averaging over model space
formed by all relevant covariates. An empirical approach is
suggested for specifying prior model probabilities. We
demonstrated through simulated microarray data that this approach
resulted in improved performance in DE gene identification
compared to the single model approaches. The flexibility of this
approach is demonstrated through our analysis of data from two
observational microarray studies.
\end{abstract}

% KEYWORDS

\end{frontmatter}

%s1 ###
\section{Introduction}\label{sec1}
In recent years, as the rapid advances in biotechnology have markedly
driven down the cost of microarray experiments, more and more large scale
studies are carried out with heterogeneous samples, conveniently
collected from subjects of different phenotypic characteristics and
exposure histories. Such microarray studies are considered
observational rather than experimental in
nature [\citet{Potter2003Epidemiology}] because the
  effects of confounding or
correlation in covariates need to be properly handled. The sample
complexity of such studies presents both opportunities and challenges
to the analysis. Considering the differential gene
expression studies, with multifaceted sample
characteristics, one may explore more complex questions that
are not possible with a more homogeneous sample such as
  the identification of differentially expressed (DE) genes
associated with not just one sample characteristic but multiple
characteristics and/or their interactions. For example,
  \citet{BoyleGumusDannenberg2010} investigated DE genes associated
  with smoking as well as smoking $\times$ gender interaction. In
another study involving smokers and never
smokers [\citet{Carolan2008Decreased}], microarray data were obtained
for an unbalanced lung airway epithelium sample involving different
tissue sites from subjects of different gender, age and ethnicity. An
interesting question is to identify DE genes associated with either a
single or multiple sample characteristics. To address
  these questions, one needs to quantify the strength of association
  between the expression of each gene and a set of sample characteristics. This
  differs from the gene set enrichment analysis [\citet{Efron2007};
    \citet{Efron2010}], where the interest is to quantify the strength of
  association between a set of genes and a single sample
  characteristic. Direct application of currently available
approaches to these questions does not provide a coherent solution and
has clear limitations.

Methods for identifying DE genes are typically based on the ranking of
statistics for between group differences associated with one sample
characteristic (also known as a factor or a covariate), such as the $t$-,
$F$-statistics, their nonparametric counterparts, their modified forms,
or the Bayesian versions [see \citet{Jeffery2006Comp} for an excellent
review of the various approaches]. These methods are suited for well
controlled experiments. Their lack of control for confounding factors attracts
increasing concern when applied to observational microarray studies
[\citet{Potter2003Epidemiology}; \citet{Webb2007Microarrays}; \citet{Troester2009Microarrays}].
With observational samples, the results may be
confounded by a variety of sample characteristics, such as age, sex,
genetic profile, exposure and treatment history, etc., which can lead
to an increased number of false discoveries.
Recent studies by \citet{Scheid2007Comp} and \citet{Leek2007PLoS} suggested
that hidden traces of unknown confounders may exist in DE gene studies and
that ranking statistics need to be adjusted accordingly. To account
for the effects of possible confounders, several approaches have been
adapted from traditional observational studies and applied to microarray
data [\citet{Smyth2004Linear}; \citet{Hummel2008Global}],
including adjustment via multiple regression
  on known confounders or on surrogate variables for unknown
  confounders [\citet{Leek2007PLoS}], or via a matched study design
  [\citet{Heller2009Matching}].

Regardless of covariate adjustment, the aforementioned approaches
rank the genes based on the effect sizes estimated using the same
model, that is, a model with the same structure and same
  set of covariates, for all genes. Such a single model approach can
be problematic for high-dimensional microarray data because
different genes may be involved in different biological processes and
their expression may be affected by different sets of covariates. More
specifically, as shown in Section~\ref{sec2}, such an approach leads to model
misspecification for a certain proportion of the genes and does not
offer the same level of accuracy and efficiency for the effect size
estimation for genes under investigation.

To avoid model misspecification in microarray data analysis, an ideal
solution could be to apply different models to different sets of genes
whereby each model contains only the set of covariates relevant to the
genes it is describing. Identifying appropriate models for
different sets of genes can be challenging because model uncertainty
makes it difficult to identify a single best model. The Bayesian model
averaging (BMA) approach offers an attractive alternative solution to
this problem. \citet{HoetingStatSci1999} provides a review of this approach in
more traditional settings. In recent years, BMA approaches have been
developed to handle various problems involving high throughput genetic
data. For example, they were used to improve the
  assessment of candidate gene effects in the genome-wide
  association studies [\citet{Wu2010Genetica}; \citet{Xu2011AOAS}] and to
  improve sample classification using gene expression microarray
  data [\citet{Yeung2005Bioinfo}]. They have also been shown to improve
  the DE gene detection in settings where the microarray data involved two
  different distributional assumptions [\citet{Sebastiani2006}] or were
  from different sources [\citet{Conlon2006}]. All these approaches
are computationally expensive, as MCMC simulation is
  used to obtain estimates of model parameters. In this study, we
propose a BMA
approach for observational microarray studies based on linear
regression models. It does not require MCMC simulations for
estimating model parameters and offers a flexible and coherent
framework to identify simultaneously DE genes associated with a
single factor, multiple factors and/or their interactions.

In the next section we discuss limitations of single
model approaches. In particular, we evaluate the impact of model
misspecification from such approaches on DE gene finding. We also
identify parameters that affect the magnitude of impact. In Section~\ref{sec3}
we propose to find DE genes with a~BMA approach that properly
  controls for sample heterogeneity and model uncertainty. In
Section~\ref{sec4} we compare the performances of ranking statistics
based on a simple model, a complex model and the BMA approach in
simulated microarray studies. Section~\ref{sec5} concludes with
  applications of BMA to two existing microarray data
sets. Our analysis supports the utility of the BMA
  method as a useful tool for capturing and quantifying the complex
  relationship between gene expression patterns and sample
  characteristics in observational microarray studies.

%
%s2 ###
\section{Limitation of the single model approaches}\label{sec2}
\label{section2}
In this section we consider a general framework to describe
gene expression variations in microarrays. Under this framework,
we argue that the single model approaches to DE gene detection are
overly simplified and subjected to the impact of model
misspecification, for example, the omission of relevant
covariates when a~simple model is used and the inclusion of
irrelevant covariates when a complex model is used. The consequences
of such model misspecification have been discussed
extensively in the linear regression
setting [\citeauthor{RaoPS1971} (\citeyear{RaoPS1971,RaoPS1973}); \citet{RosenbergLevy}]. The
implication of these results, however, has not been fully investigated
in DE gene studies. In this section we evaluate the consequences of
model misspecification from the single model approaches on performance
measures often used in DE gene studies, including the false discovery
rate (FDR) and sensitivity. We conclude this section with a summary
of the main results.

%
%s2.1 ###
\subsection{Notation}\label{sec2.1}
We consider an observational microarray study which aims to identify
DE genes associated with different values of a factor $X_{1}$, for
example, cigarette smoking exposure. Expression profiles of $J$ genes are
obtained for $n$ subjects with different values of $X_{1}$. Without
loss of generality, a typical model for identifying $X_{1}$ related DE
genes can be written as
%e2.1 ###
\begin{equation}
y_{ij}=\beta_{0j}+\beta_{1j}x_{1i}+\cdots +\beta_{kj}x_{ki}+\eta_{ij}
\label{model1}
\end{equation}
or
%e2.2 ###
\begin{equation}
y_{ij}=\alpha_{0j}+\alpha_{1j}x_{1i}+\cdots +\alpha_{kj}x_{ki}+
\alpha_{(k+1)j}x_{(k+1)i}+\epsilon_{ij},
\label{model2}
\end{equation}
where $y_{ij}$ is the normalized and typically log-transformed
expression level of gene $j$ in subject $i$; $x_{1i}$ is the factor
level for $X_1$ in subject $i$; $x_{2i},\ldots, x_{ki}$ are
levels for other factors, denoted by $X_2,\ldots, X_k$, that affect
the expression of all the genes, for example, experimental parameters
involved in the microarray experiments; $x_{(k+1)i}$ is the level
of a potential confounding factor $X_{k+1}$, for example, gender, age, race,
alcohol exposure, etc.; $\eta_{ij}$ and $\epsilon_{ij}$ denote normally
distributed random errors.

To identify DE genes related to $X_{1}$, $p$-values based on $t$-statistic of
estimate of either $\beta_{1j}$ or $\alpha_{1j}$ can be used as the
ranking statistics. If model (\ref{model1}) is used, the relevant
$t$-statistic for gene $j$ is
$t_{M_1,1j}=\hat{\beta}_{1j}/sd(\hat{\beta}_{1j})$, where
$\hat{\beta}_{1j}$ is the least square estimate of $\beta_{1j}$. If
model (\ref{model2}) is used, the $t$-statistic for gene $j$ is
calculated as\vadjust{\goodbreak}
$t_{M_2,1j}=\hat{\alpha}_{1j}/sd(\hat{\alpha}_{1j})$. It can be shown
that the two statistics are related as follows:
%e2.3 ###
\begin{equation}
  t_{M_1,1j}=\frac{S_{1\cdot 23\cdots k}}{S_{1\cdot 23\cdots k+1}}
  t_{M_2,1j}+ \frac{S^{-2}_{k+1 \cdot 1\cdots k} b_{k+1,1} e_{k+1}^{T}
  Y_j}{sd(\hat{\beta}_{1j})},
\end{equation}
where $S^{2}_{k+1 \cdot 1\cdots k}$, $b_{k+1}$ and $e_{k+1}$ are
the residual sum of squares, least square parameter estimates and
residual, respectively, from the following auxiliary regression equation:
%e2.4 ###
\begin{equation}\label{aux}
X_{k+1}=Xb_{k+1}+e_{k+1},
\end{equation}
where $X=(X_1,\ldots, X_{k})$. $S^2_{1\cdot 23\cdots k.k+1}$
is the residual sum of squares for the auxiliary regression with $X_1$
as the outcome and $X_2,\ldots, X_{k+1}$ as the covariates.

For an observational microarray study, such single model approach with
or without covariate adjustment has an intrinsic
limitation, that is, neither model can be the true
model for all the genes. For the aforementioned hypothetical
microarray study, model (\ref{model1}) is
the true model only for genes not related to $X_{k+1}$ ($X_{k+1}$ null
genes, or $M_1$ genes), and model (\ref{model2}) is the true model
only for genes related to $X_{k+1}$ ($X_{k+1}$ DE genes, or $M_2$
genes). Based on these considerations, a multi-model approach that uses
$p$-values of $t_{M_1,1\cdot}$ to rank the $M_1$ genes and $p$-values of
$t_{M_2,1\cdot}$ to rank the $M_2$ genes is preferable.

The performance difference between the single model
and the multi-model approaches can be compared by utilizing the
relationship between the two $t$-statistics. Let $F_{1}(t)$ and
$F_{2}(t)$ be the density distributions of the ranking statistics
$t_{M_1,1\cdot}$ and $t_{M_2,1\cdot}$, respectively. Under
the multi-model approach, the density distribution of the ranking
statistics can be written as
\[
F(t)=(1-f)F_1(t)+f F_2(t),
\]
where $f$ is the proportion of $M_2$ genes. $F_1(t)$ and $F_2(t)$ can
further be written as
\begin{eqnarray*}
  F_1(t)&=&(1-p_{1})F_{10}(t)+p_1 F_{11}(t),\\
  F_2(t)&=&(1-p_{2})F_{20}(t)+p_2 F_{21}(t),
\end{eqnarray*}
where $p_1$ and $p_2$ are the proportions of DE genes in $M_1$ and $M_2$
genes, $F_{\cdot 0}(t)$ and $F_{\cdot 1}(t)$ are
distributions of the test statistic for the null and DE genes,
respectively. For a given cutoff $c>0$, the false discovery rate and
sensitivity can be calculated as
%e2.5 ###
\begin{eqnarray}\label{FDRformula}
  \mathit{FDR}(c)&=&\frac{(1-f)(1-p_1)[1-F_{10}(c)]}
  {(1-f)[1-F_1(c)]+ f[1-F_2(c)]}
 \nonumber
 \\[-8pt]
 \\[-8pt]
  &&{}+\frac{f(1-p_2)[1-F_{20}(c)]}{(1-f)[1-F_1(c)]+ f[1-F_2(c)]}
\nonumber
\end{eqnarray}
and
\[
S(c)= 2(1-f)p_1[1-F_{11}(c)]+2fp_2[1-F_{21}(c)].\vadjust{\goodbreak}
\]
We discuss the impact of the two single model approaches on the FDR
and sensitivity separately.

%
%s2.2 ###
\subsection{Single model without covariate adjustment}\label{sec2.2}
When model (\ref{model1}) is used, the FDR can be written as
\begin{eqnarray*}
\mathit{FDR}^{M_1}(c)&=&\frac{(1-f)(1-p_1)[1-F_{10}(c)]}
{(1-f)[1-F_1(c)]+ f[1-F^{M_1}_2(c)]}\\
&&{}+\frac{f(1-p_2)[1-F^{M_1}_{20}(c)]}
{(1-f)[1-F_1(c)]+ f[1-F^{M_1}_2(c)]}.
\end{eqnarray*}
 The sensitivity can be written as
\[
S^{M_1}(c)= 2(1-f)p_1[1-F_{11}(c)]+2fp_2[1-F^{M_1}_{21}(c)].
\]
Superscript $M_1$ is used to denote that the distribution of
$t$-statistic is derived from model (\ref{model1}), which is misspecified
for the $M_2$ genes because of omitting relevant covariate $X_{k+1}$.

Omission of relevant covariate leads to bias in the model parameter
estimates [\citet{RaoPS1971}]. Specifically, the bias can be written as
%e2.6 ###
\begin{equation}
\label{eqbias}
\mathit{Bias}(\hat{\beta}_{1j})=E(S^{-2}_{k+1 \cdot 1\cdots k}b_{k+1,1}e_{k+1}^{T}Y_j)
                             =\alpha_{k+1,j}\cdot b_{k+1,1},
\end{equation}
where $b_{k+1,1.23\cdots k}$ is the least square estimate of the
parameter associated with $X_1$ in the auxiliary regression
 (\ref{aux}). Therefore, we have for the $M_2$ gene~$j$
\[
E(t_{M_1,1j})\approx \frac{S_{1\cdot 23\cdots k}}{S_{1\cdot
 23\cdots k+1}}
  \biggl[E(t_{M_2,1j})+\frac{b_{k+1,1}\alpha_{k+1}}{\sigma_{2j}/S_{1\cdot
 23\cdots k+1}} \biggr].
\]
It is known that $
S^2_{1\cdot 23\cdots k.k+1} \leq S^2_{1\cdot 23\cdots k}$.

For the $M_2$ DE genes, because $t_{M_1,1j}$ can be greater
or less than $t_{M2,1j}$ depending on the values of $\alpha_1$ and
$\mathit{Bias}(\hat{\beta}_{1})$, $F^{M_1}_{21}(t)$ is unlikely to
be systematically different from $F_{21}(t)$ and results in great
changes in sensitivity.

However, for the $M_2$ null genes, the above results indicate
$E|t_{M_1,1j}|\geq E|t_{M_2,1j}|$, that is, the distribution of
$t_{M_2,1j}$ for the $M_2$ null genes moves away from zero. Hence,
$1-F^{M_1}_{20}(c) \geq 1-F_{20}(c)$. Let $a$ and $b$ be the
denominator and numerator of $\mathit{FDR}(c)$ as written in (\ref{FDRformula}),
respectively. Let $\delta$ be the difference between the numerators of
$\mathit{FDR}^{M_1}(c)$ and $\mathit{FDR}(c)$, that is,
\[
\delta=f(1-p_2)\{[1-F^{M_1}_{20}(c)]-[1-F_{20}(c)]\},
\]
and $\delta'$ be the difference between the denominators of the two
FDRs,
\begin{eqnarray*}
\delta'&=&f(1-p_2)\{[1-F^{M_1}_{20}(c)]-[1-F_{20}(c)]\}\\
        &&{}+fp_2\{[1-F^{M_1}_{21}(c)]-[1-F_{21}(c)]\}.
\end{eqnarray*}
As discussed above, $[1-F^{M_1}_{21}(c)]$ is comparable to
$[1-F_{21}(c)]$ because the bias is unlikely to lead to systematic
difference between\vadjust{\goodbreak} $F^{M_1}_{21}(t)$ and $F_{21}(t)$. Additionally,
$p_2$ generally is much smaller than $1-p_2$ in
microarrays. Therefore, $\delta'\approx \delta$ and $\mathit{FDR}^{M_1}(c)$ can
be approximated by $(b+\delta)/(a+\delta)$. Since
$(b+\delta)/(a+\delta) \geq b/a$ for any $a>b>0$ and $\delta\geq 0$,
this indicates $\mathit{FDR}^{M_1}(c) \geq \mathit{FDR}(c)$, that is, increased FDR with
this single model approach.

%
%s2.3 ###
\subsection{Single model with covariate adjustment}\label{sec2.3}
When model (\ref{model2}) is used, the FDR and
sensitivity at a given cutoff can be written as
\begin{eqnarray*}
\mathit{FDR}^{M_2}(c)&=& \frac{(1-f)(1-p_1)[1-F^{M_2}_{10}(c)]}
{(1-f)[1-F^{M_2}_1(c)]+ f[1-F_2(c)]}\\
&&{}+     \frac{f(1-p_2)[1-F_{20}(c)]}
{(1-f)[1-F^{M_2}_1(c)]+ f[1-F_2(c)]}
\end{eqnarray*}
 and
\[
S^{M_2}(c)= 2(1-f)p_1[1-F^{M_2}_{11}(c)]+2fp_2[1-F_{21}(c)],
\]
due to the potential change in the distributions of test
statistics for the $M_1$ genes. The relationship of the
two $t$-statistics can be written as
\[
t_{M_2,1j}=\frac{S_{1\cdot 23\cdots k+1}}{S_{1\cdot 23\cdots k}} t_{M_1,1j}+
   \frac{S^{-2}_{k+1 \cdot 1\cdots k}b_{k+1,1}e_{k+1}^{T}Y_j}{sd(\hat{\alpha}_{1j})}.
\]
It is known that, with the inclusion of an irrelevant covariate,
model (\ref{model2}) does not result in a biased parameter estimate for
the $M_{1}$ genes. However, since $sd(\hat{\beta}_{1j}) \leq
sd(\hat{\alpha}_{1j})$ in general, $E(|t_{M_1,1}|) \geq
E(|t_{M_2,1}|)$ for $M_1$ DE genes. Therefore, the distribution
$F^{M_2}_{11}(t)$ moves toward 0 and results in $S^{M_2}(c)\leq S(c)$,
that is, reduced sensitivity in detecting DE genes in $M_1$ genes. As
$|t_{M_1,1}|$ in general is likely to be greater than $|t_{M_2,1}|$,
$F^{M2}_{10}$ also shrinks toward 0. It is likely that $\mathit{FDR}^{M_2}(c)$
will be comparable to $\mathit{FDR}^{M_1}(c)$. Hence, reduced sensitivity in
detecting DE genes in $M_1$ genes will be the main consequence
resulted from applying the complex model for all the genes.

%
%s2.4 ###
\subsection{Summary}\label{sec2.4}
The above results suggested that the single model approaches with or
without covariate adjustment can lead to inferior performance. It is
expected that the impact on FDR and sensitivity could be greater if
more $X_{k+1}$-like covariates exist in the sample. These results
will be further demonstrated in the simulation study. The above
discussion also suggested that the performance for DE gene detection
can be improved by applying the correct model for the right sets of
genes. Yet, such knowledge is commonly not available beforehand. In the
following section, we propose a BMA approach as a practical substitute
for the multi-model approach for DE gene detection that
takes into account both sample heterogeneity and model uncertainty.

%
%s3 ###
\section{A Bayesian model averaging approach}\label{sec3}
In this section we discuss an efficient Bayesian model averaging
approach to identifying DE genes associated with a covariate of
interest. The methodology proposed in this paper is
  closely related to methods discussed in \citet{LiangJASA2008} and we
  largely follow their notation. Consider a series of possible models for
describing the expression pattern of each gene. Let
$\boldsymbol{\gamma}= (\gamma_{1},\ldots, \gamma_{K})$ be a binary
vector of length $K$, with each element indicating the inclusion
status of the $k$th covariate in the model, that is,
\[
\gamma_{k}=\cases{
   0, &\quad if $\beta_{k}=0$,  \cr
   1, &\quad if $\beta_{k}\ne 0$.}
\]
Each model in the model space can then be labeled by $\boldsymbol{\gamma}$,
namely, $\mathcal{M}_{\boldsymbol{\gamma}}$. For gene $j$, $j = 1, \ldots,
J$, the model can be written as
\[
  \mathcal{M}_{\boldsymbol{\gamma}j}\dvtx
  \mathbf{Y}_j = \alpha_{\boldsymbol{\gamma}j} {\bf 1}_{n} +
  \mathbf{X}_{\boldsymbol{\gamma}}
  \boldsymbol{\beta}_{\boldsymbol{\gamma}j} +
  \mbox{N}({\bf 0}, \phi_{\boldsymbol{\gamma}j}^{-1}\n I_n),
\]
 where $\alpha_{\boldsymbol{\gamma}j}$ is the intercept term;
$\mathbf{X}_{\boldsymbol{\gamma}}$ is the submatrix of
$\mathbf{X}$ consisting of columns associated with nonzero $\gamma_{k}$;
$\boldsymbol{\beta}_{\boldsymbol{\gamma}j}$ and
$\phi_{\boldsymbol{\gamma}j}$ are parameters under this model.

The marginal posterior inclusion probability for variable $X_k$ and
gene $j$, is then defined as
%e3.1 ###
\begin{equation}
  P_{kj}=P(\gamma_{kj}\neq0|\mathbf{Y}_j) =
  \sum_{\boldsymbol{\gamma}}{{\bf 1}_{\gamma_{kj}=1}\times
    P(\mathcal{M}_{\boldsymbol{\gamma}j}|\mathbf{Y}_j)},
  \label{postvarinclprob}
\end{equation}
which is the sum of posterior probabilities of all models that
include the covariate of interest. It quantifies the strength of
association between covariate~$X_k$ and the expression level of
the $j$th gene and can be used to rank the DE genes.

The posterior model probability for $\mathcal{M}_{\boldsymbol{\gamma}j}$
can be calculated based on Bayes factors of pairs of models, for example,
%e3.2 ###
\begin{equation}
\label{postmodelprob}
P(\mathcal{M}_{\boldsymbol{\gamma}j}|\mathbf{Y}_j) =
\frac{p(\mathcal{M}_{\boldsymbol{\gamma}j})
  \mathit{BF}(\mathcal{M}_{\boldsymbol{\gamma}j}\dvtx\mathcal{M}_{\mathbf{0}j})}
     {\sum _{\boldsymbol{\gamma}^{\prime}}
       p(\mathcal{M}_{\boldsymbol{\gamma}^{\prime}j})
       \mathit{BF}(\mathcal{M}_{\boldsymbol{\gamma}^{\prime}{j}}\dvtx
       \mathcal{M}_{\mathbf{0}j})},
\end{equation}
where $p(\mathcal{M}_{\boldsymbol{\gamma}j})$ is the
prior model probability for genes measured in the microarray
experiment and the Bayes
factor $\mathit{BF}(\mathcal{M}_{\boldsymbol{\gamma}j}\dvtx\mathcal{M}_{\mathbf{0}j})$
is defined as
\[
\mathit{BF}(\mathcal{M}_{\boldsymbol{\gamma}j}\dvtx\mathcal{M}_{\mathbf{0}j}) =
\frac{f(\mathbf{Y}_j|\mathcal{M}_{\boldsymbol{\gamma}j})}
{f(\mathbf{Y}_j|\mathcal{M}_{\mathbf{0}j})},
\]
that is, the ratio of marginal likelihood under
$\mathcal{M}_{\boldsymbol{\gamma}j}$ and the base model,~$\mathcal{M}_{\mathbf{0}j}$. Here the null model (i.e., the model
with only the intercept term) is used as the base model. For
$\mathcal{M}_{\boldsymbol{\gamma}j}$, the marginal likelihood is obtained
by integrating out the model parameters from the joint posterior
probability
\[
f(\mathbf{Y}_j|\mathcal{M}_{\boldsymbol{\gamma}j})=\int
f(\mathbf{Y}_j|\boldsymbol{\Theta}_{\boldsymbol{\gamma}j})
\pi(\boldsymbol{\Theta}_{\boldsymbol{\gamma}j}| \mathcal{M}_{\boldsymbol{\gamma}j})
\, d\boldsymbol{\Theta}_{\boldsymbol{\gamma}j},
\]
where $\boldsymbol{\Theta}_{\boldsymbol{\gamma}j}=(\alpha_{\boldsymbol{\gamma}j},
\boldsymbol{\beta}_{\boldsymbol{\gamma}j}, \phi_{\boldsymbol{\gamma}j})$, and
$\pi(\boldsymbol{\Theta}_{\boldsymbol{\gamma}j}
|\mathcal{M}_{\boldsymbol{\gamma}j})$ is the
prior of model parameters under $\mathcal{M}_{\boldsymbol{\gamma}j}$.\vadjust{\goodbreak}

To determine the Bayes factor, proper priors,
  $\pi(\boldsymbol{\Theta}_{\boldsymbol{\gamma}j}|\mathcal{M}_{\boldsymbol{\gamma}j})$, are
needed. We utilized the Zellner--Siow prior for model
parameters [\citet{ZellnerSiow1980}] in our study. \citet{LiangJASA2008} have
shown that this prior resolves several consistency issues
associated with fixed $g$-priors while retaining several attractive
properties such as adaptivity, good shrinkage properties, robustness
  to the misspecification of $g$
and fast marginal likelihood calculation. When comparing two nested
models as in our case, a flat prior is placed on
common coefficients, ($\alpha_{\boldsymbol{\gamma}j}$,
$\phi_{\boldsymbol{\gamma}j}$), where $\pi(\alpha_{\boldsymbol{\gamma}j},
\phi_{\boldsymbol{\gamma}j}|\mathcal{M}_{\boldsymbol{\gamma}j}) \propto 1 /
\phi_{\boldsymbol{\gamma} j}$, and a Cauchy prior on the remaining
parameters, $\boldsymbol{\beta}_{\boldsymbol{\gamma}j}$. The
multivariate Cauchy prior can then be represented as a mixture of
$g$-priors with an Inv-gamma($1/2$, $n/2$) prior on $g$, that is,
\[
\pi(\boldsymbol{\beta}_{\boldsymbol{\gamma} j}|\phi_{\boldsymbol{\gamma} j},
\mathcal{M}_{\boldsymbol{\gamma}j}) \propto
\int N \biggl(\boldsymbol{\beta}_{\boldsymbol{\gamma} j}|\mathbf{0},
\frac{g}{\phi_{\boldsymbol{\gamma} j}}(\mathbf{X}_{\boldsymbol{\gamma}}^T
\mathbf{X}_{\boldsymbol{\gamma}})^{-1} \biggr)\pi(g)\,dg,
\]
with
\[
\pi(g)=\frac{\sqrt{n/2}}{\Gamma(1/2)}g^{-3/2}e^{-n/(2g)}.
\]
The Bayes factor in equation (\ref{postmodelprob})
can be written in closed form as
\begin{eqnarray*}
\mathit{BF}(\mathcal{M}_{\boldsymbol{\gamma}j}\dvtx \mathcal{M}_{\mathbf{0}j})&=&
\int_0^{\infty}(1+g)^{(n-1-\rho_{\boldsymbol{\gamma}j})/2}\\
 & &\hphantom{\int_0^{\infty}}{}\times[1+(1-R_{\boldsymbol{\gamma}j}^2)g]^{-(n-1)/2}\pi(g)\, {d} g,
\end{eqnarray*}
where $\rho_{\boldsymbol{\gamma}j}$ denotes the number of covariates
included in $\mathcal{M}_{\boldsymbol{\gamma}j}$ and
$R_{\boldsymbol{\gamma}j}^2$ is the ordinary coefficient of
determination of this model. This quantity can be obtained through
direct numerical integration or through the Laplace approximation.

In addition to the prior
  $\pi(\boldsymbol{\Theta}_{\boldsymbol{\gamma}j}|\mathcal{M}_{\boldsymbol{\gamma}j})$ on
  model parameters, one must also choose a prior on the models
  themselves, which relates directly to multiplicity.
  \citet{ScottAOS2010} discussed several prior model
  probability choices regarding their effects on multiplicity-control
  for multiple models in a~conventional Bayesian model
  selection/averaging setting involving one outcome variable. With the
  high throughput data, typically,
the prior model probabilities should reflect our prior belief about
the distribution of the models among the genes in the
transcriptome, which can be difficult to quantify. A
uniform prior assumed equal probabilities of the models
can be problematic when thousands of genes are evaluated
simultaneously because it puts an unrealistically low weight to the
null model. When the resulting posterior model probabilities are used
to estimate the posterior expected FDR (\textit{pe}FDR) [\citet{NewtonBiostat2004}], great
underestimation can occur [\citet{SartorBMCBioInfo2006};
  \citet{Cao2009}]. Correctly estimating FDR under the Bayesian framework
remains an active research field [\citet{Efron2008}]. Recent discussions
and attempts have largely been focused on statistics derived from\vadjust{\goodbreak}
single model approaches [\citet{MullerGP2007}; \citet{Cao2010}]. In our case,
proper control for multiplicity derived from multiple genes
and multiple models becomes even more challenging.

We believe that the prior should lead to a reasonably well calibrated
posterior model probability that measures the model's ability for
describing the data. We propose an empirical approach
to obtain estimates for the prior model probabilities,
$p(\mathcal{M}_{\boldsymbol{\gamma}j})$, under the assumption that the
prior probabilities of a given model are the same across genes, that is,
$p(\mathcal{M}_{\boldsymbol{\gamma}j})=p(\mathcal{M}_{\boldsymbol{\gamma}})$.
We first estimate the proportion of DE genes described by a nonnull
model~${\boldsymbol{\gamma}}$, $\omega_{\boldsymbol{\gamma}}$, using Bayes
factors. Since $\mathit{BF}(\mathcal{M}_{\boldsymbol{\gamma}}\dvtx
\mathcal{M}_{\mathbf{0}})>c$, $c \ge 1$ suggests evidence
against the null model [\citet{KassRaftery1995}], we can estimate
$\omega_{\boldsymbol{\gamma}}$ as follows:
\[
  \omega_{\boldsymbol{\gamma}}=\frac{1}{J} \sum_{j}
        {\bf 1}_{[\mathit{BF}(\mathcal{M}_{\boldsymbol{\gamma}j}\dvtx
  \mathcal{M}_{\mathbf{0}j})= \max(\mathit{BF}_{j})]}\cdot
        {\bf 1}_{[\mathit{BF}(\mathcal{M}_{\boldsymbol{\gamma}j}\dvtx
          \mathcal{M}_{\mathbf{0}j})>c]},
\]
where $\mathit{BF}_{j}$ is a vector of null-based Bayes factors
for gene $j$. Therefore, $\omega_{\boldsymbol{\gamma}}$ represents the
proportion of genes for which model ${\boldsymbol{\gamma}}$ is the best
model in terms of Bayes factors.
Given that Bayes factors based on the Zellner--Siow prior are consistent
for model selection whether or not the true model is
null [\citet{LiangJASA2008}], this estimator is a consistent estimator
of the proportion of genes expressing in a pattern specified by the
model. In our simulation studies, we found that
fixing $c$ at 1 resulted in $\omega_{\boldsymbol{\gamma}}$ being
close to the truth in most settings. Second, we argue that if the
prior model probabilities,
$p(\mathcal{M}_{\boldsymbol{\gamma}})$, result in the
equality between the overall $pe$FDR under
$\mathcal{M}_{\boldsymbol{\gamma}}$ and $1-\omega_{\boldsymbol{\gamma}}$,
reasonable calibration of the posterior model probabilities can be
achieved. This suggests the following relationship between
$p(\mathcal{M}_{\boldsymbol{\gamma}})$ and $\omega_{\boldsymbol{\gamma}}$, that is,
\[
\omega_{\boldsymbol{\gamma}}=\frac{1}{J}\sum_{j}
\frac{\mathit{BF}(\mathcal{M}_{\boldsymbol{\gamma}j}\dvtx
  \mathcal{M}_{\mathbf{0}j})p(\mathcal{M}_{\boldsymbol{\gamma}})}
{\sum_{\boldsymbol{\gamma'}}\mathit{BF}(\mathcal{M}_{\boldsymbol{\gamma}'j}\dvtx
\mathcal{M}_{\mathbf{0}j}) p(\mathcal{M}_{\boldsymbol{\gamma}'})}.
\]
Hence, $p(\mathcal{M}_{\boldsymbol{\gamma}})$ can be
obtained by iteratively updating the following equation:
  \[
  p^{(l)}(\mathcal{M}_{\boldsymbol{\gamma}})=\frac{\sum_{j}
    {\bf 1}_{[\mathit{BF}(\mathcal{M}_{\boldsymbol{\gamma}j}\dvtx
        \mathcal{M}_{\mathbf{0}j})= \max(\mathit{BF}_{j})]}\cdot
    {\bf 1}_{[\mathit{BF}(\mathcal{M}_{\boldsymbol{\gamma}j}\dvtx
        \mathcal{M}_{\mathbf{0}j})>c]}}{
    \sum_{j}[\mathit{BF}(\mathcal{M}_{\boldsymbol{\gamma}j}\dvtx \mathcal{M}_{\mathbf{0}j})/
    \sum_{\boldsymbol{\gamma}'}\mathit{BF}(\mathcal{M}_{\boldsymbol{\gamma}'j}\dvtx
    \mathcal{M}_{\mathbf{0}j}) p^{(l-1)}(\mathcal{M}_{\boldsymbol{\gamma}'})]}
  \]
  under the constraint $\sum_{\boldsymbol{\gamma}}
  p^{(l)}(\mathcal{M}_{\boldsymbol{\gamma}})=1$, where $l$ denotes the
  iteration step. In our experience, 30 iterations were adequate to
  result in convergence.
At present stage, theoretical justification for this prior choice for
multiplicity control is still lacking. We resort to the simulation
study to show that this prior choice  led to improved performance in
both the ranking of the genes and in direct FDR estimation compared
with the uniform prior.

%
%s4 ###
\section{Simulation study}\label{sec4}
Simulation studies were designed to compare the single
  model approaches with and without covariate adjustment and the ``gold
  standard'' multiple-model approach with the correct covariate\vadjust{\goodbreak}
  adjustments, as well as the performance of BMA over single-model
  approaches when a multi-model approach is appropriate. Bias and
  efficiency in each approach and sensitivity to the choice of prior
  on the set of models also will be discussed.

%s4.1 ###
\subsection{Simulation of microarray data}\label{sec4.1}
The microarray data were simulated to mimic an observational
study for identifying genes associated with a binary variable, for
example, the smoking status ($s$), in a sample with two
potential confounders, gender ($g$) and
heavy alcohol drinking ($d$) which are also
binary. A detailed data generation scheme for the subject
characteristics is provided in \citet{Zhou2011}. Marginally, half of the subjects are
assumed to be females, smokers or heavy drinkers. We also assume complex
correlation among these covariates. First, $s$ is correlated with
both $g$ and $d$. Specifically, in smokers, 75\% are
males and 80\% are heavy drinkers; while in nonsmokers, 25\% are
males and 20\% are heavy drinkers. Second, $g$ is also correlated
with $d$. Specifically, 75\% of male subjects are heavy drinkers,
while 25\% of females are heavy drinkers. Proportions of subjects in
groups defined by categories of the covariates are provided in \citet{Zhou2011}. Each microarray data set
consists of the expression of 10\mbox{,}000 genes from $n$ subjects. Gene
expression for each subject was simulated based on the following model:
\[
y_{ij}=\beta_{1j}s_{i}+\beta_{2j}g_{i}+\beta_{3j}d_{i}+\varepsilon_{ij},
\]
where $\beta_{.j}$ takes either 0 or nonzero values generated from
normal distributions with variances generated following procedures
similar to that described by \citet{Smyth2004Linear}. Detailed
procedures for generating the simulated microarray data are provided
in \citet{Zhou2011}. Each simulation
setting was characterized by values of the following parameters:
$f_s$, $f_g$ and $f_d$, the proportion of genes affected
by smoking ($s$), gender ($g$) or heavy drinking ($d$), respectively,
and $n$, the sample size. Both moderate and relatively large sample sizes were
considered, $n=40$ and $n=80$. For each setting, we simulated $10$
microarray data sets. The reported results were averaged over the
results obtained for each data set.

%{\renewcommand{\baselinestretch}{1}
%t1 ###
\begin{table}
\tabcolsep=0pt
    \caption{False discovery rate ($\mathit{FDR}$) and
      sensitivity ($S$), in \%,
      among the top smoking related genes identified with a $p$-value
      cutoff of 0.001 using ranking statistics based on the single model
      approach without covariate adjustment (${\mathit{SM}}_1$), the single model
      approach with covariate adjustment ($\mathit{SM}_2$), the surrogate
      variable analysis approach ($\mathit{SVA}$) and the ``gold standard'' multi-model
      approach ($\mathit{MM}$). FDR and sensitivity arising from $g0d0$ genes
      (i.e., genes not associated with $d$ and $g$) were
      included. Microarray data sets were simulated based on
      various settings defined by proportion of genes associated with
      each covariate: $f_s$, $f_g$, $f_d$, and the sample size $n$}
    \label{tablefdr}
\begin{tabular*}{\textwidth}{@{\extracolsep{4in minus 4in}}lcccccccc@{}}
      \hline
      \textbf{Methods}
      & \multicolumn{4}{c}{$\boldsymbol{n=40}$}&\multicolumn{4}{c@{}}{$\boldsymbol{n=80}$}\\[-5pt]
 & \multicolumn{4}{c}{\hrulefill}&\multicolumn{4}{c@{}}{\hrulefill}\\
     &$\boldsymbol{\mathit{FDR}_{g0d0}}$&$\boldsymbol{\mathit{FDR}_{total}}$&$\boldsymbol{S_{g0d0}}$&
      $\boldsymbol{S_{total}}$&$\boldsymbol{\mathit{FDR}_{g0d0}}$&$\boldsymbol{\mathit{FDR}_{total}}$&$\boldsymbol{S_{g0d0}}$&$\boldsymbol{S_{total}}$\\
      \hline
      \multicolumn{9}{@{}l@{}}{$f_s=0.10$, $f_g=0.05$, $f_d=0$}\\
      ${\mathit{SM}}_1$ & \hphantom{0}$4.2$& \hphantom{0}$6.5$& $14.1$& $14.9$& $2.3$& \hphantom{0}$8.2$& $28.5$& $30.2$\\
      ${\mathit{SM}}_2$ & \hphantom{0}$6.1$& \hphantom{0}$6.5$& $10.0$& $10.4$& $2.2$& \hphantom{0}$2.5$& $23.1$& $24.3$\\
      $\mathit{SVA}$ & \hphantom{0}$6.2$& \hphantom{0}$6.7$& \hphantom{0}$9.3$& \hphantom{0}$9.7$& $2.2$& \hphantom{0}$2.3$& $22.7$& $24.0$\\
      $\mathit{MM}$     & \hphantom{0}$4.4$& \hphantom{0}$4.8$& $14.1$& $14.5$& $2.4$& \hphantom{0}$2.6$& $28.5$& $29.7$\\[5pt]
      \multicolumn{9}{@{}l@{}}{$f_s=0.05$, $f_g=0.10$, $f_d=0$}\\
      ${\mathit{SM}}_1$ & \hphantom{0}$8.5$&$18.0$& $12.9$& $14.5$& $3.6$& $22.9$& $26.6$& $29.8$\\
      ${\mathit{SM}}_2$ &$10.5$&$11.7$& \hphantom{0}$9.4$& $10.6$& $5.1$& \hphantom{0}$5.7$& $21.0$& $23.4$\\
      $\mathit{SVA}$   &$10.3$&$11.6$&\hphantom{0}$9.0$& $10.2$& $5.4$& \hphantom{0}$6.2$& $20.7$& $23.1$\\
      $\mathit{MM}$     & \hphantom{0}$9.5$& \hphantom{0}$9.5$& $12.9$& $14.1$& $4.5$& \hphantom{0}$5.0$& $26.6$& $29.0$\\[5pt]
     \multicolumn{9}{@{}l@{}}{$f_s=0.1$, $f_g=0.05$, $f_d=0.05$ }\\
      ${\mathit{SM}}_1$ &\hphantom{0}$4.1$& \hphantom{0}$8.5$& $13.4$& $15.0$& $2.6$& $12.8$& $26.4$& $29.5$\\
      ${\mathit{SM}}_2$ &\hphantom{0}$6.8$& \hphantom{0}$7.4$& \hphantom{0}$8.4$& \hphantom{0}$9.3$& $3.0$& \hphantom{0}$3.2$& $19.9$& $22.1$\\
      $\mathit{SVA}$    &\hphantom{0}$7.1$& \hphantom{0}$8.0$&\hphantom{0}$8.2$& \hphantom{0}$9.1$&$2.9$& \hphantom{0}$3.1$& $19.3$& $21.5$\\
      $\mathit{MM}$     &\hphantom{0}$4.4$& \hphantom{0}$4.9$& $13.4$& $14.3$& $3.0$& \hphantom{0}$3.1$& $26.4$& $28.8$\\[5pt]
      \multicolumn{9}{@{}l@{}}{$f_s=0.05$, $f_g=0.10$, $f_d=0.10$}\\
      ${\mathit{SM}}_1$ &\hphantom{0}$4.7$& $19.4$& $12.6$& $15.7$& $3.4$& $36.9$& $25.1$& $31.3$\\
      ${\mathit{SM}}_2$ &\hphantom{0}$9.9$& $12.6$& \hphantom{0}$7.9$& \hphantom{0}$9.6$& $4.5$& \hphantom{0}$6.1$& $18.4$& $22.6$\\
      $\mathit{SVA}$    &$10.7$& $13.1$& \hphantom{0}$7.7$& \hphantom{0}$9.4$& $4.6$& \hphantom{0}$6.0$& $18.0$& $22.4$\\
      $\mathit{MM}$     &\hphantom{0}$5.9$& \hphantom{0}$8.2$& $12.6$& $14.4$& $5.4$& \hphantom{0}$6.5$& $25.1$& $29.6$\\
      \hline
    \end{tabular*}
\end{table}

%s4.2 ###
\subsection{Performance of the single model approaches}\label{sec4.2}
In this section we compare the performances of three
single model approaches that differed by covariate adjustment, that is,
without covariate adjustment ($\mathit{SM}_1$), with adjustment to $g$ and $d$
($\mathit{SM}_2$), and with adjustment to surrogate variables of $g$ and $d$ ($\mathit{SVA}$)
[\citet{Leek2007PLoS}], and that of the gold standard
multi-model approach ($\mathit{MM}$) where the DE genes were fit with their
respective true models, that is, the adjustment for $g$ and/or
$d$ is applied only to genes truly affected by $g$ and/or $d$. The
sensitivity and FDR corresponding to the ranking statistic,
$p$-value of $s$, were obtained for each method. To show the interplay
of bias and efficiency on these performance measures, we also quantified the
contribution to these measures from genes not associated with $g$ and
$d$, denoted as $g0d0$ genes.

Table~\ref{tablefdr} shows the performance difference between the
single and multi-model approaches among top ranked genes identified
with a $p$-value cutoff of 0.001. We can see that, as discussed in
Section~\ref{section2}, ${\mathit{SM}}_1$ led to large increase in total FDR
compared to $\mathit{MM}$. The magnitude of difference increased with the
sample size, the proportion of the genes associated with the
confounder and the number of the confounders. On the other hand, the
difference in FDR contributed from  the $g0d0$ genes remained
small. Hence, the results suggested that bias in effect estimation
among genes associated with the confounders was the main cause for
the FDR increase. ${\mathit{SM}}_2$ and $\mathit{SVA}$ showed slightly greater FDR compared to
$\mathit{MM}$. This increase came mainly from $g0d0$ genes and suggested that
the effects of the efficiency loss could have a negative impact on the
total FDR, particularly in small sample size settings. A more notable
limitation of $\mathit{SM}_2$ and $\mathit{SVA}$ was the loss of sensitivity. Compared to $\mathit{MM}$,
the magnitude of sensitivity loss increased slightly with sample size
and the number of confounders.

%s4.3 ###
\subsection{Performance of the BMA approach}\label{sec4.3}
\begin{figure}
  \centering
  \begin{tabular}{cc}

\includegraphics{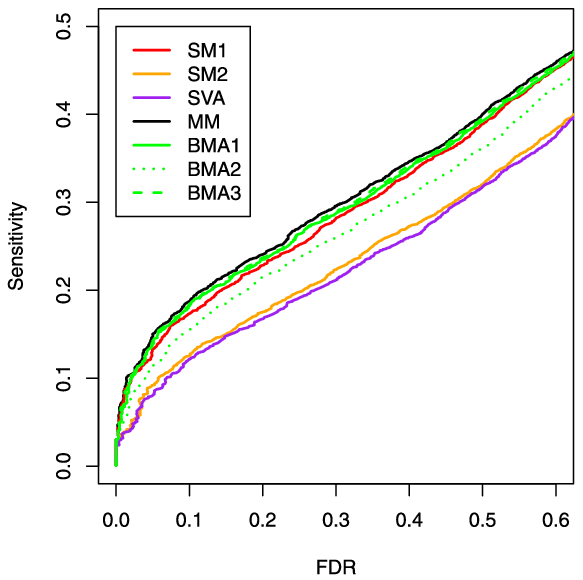}
 &

\includegraphics{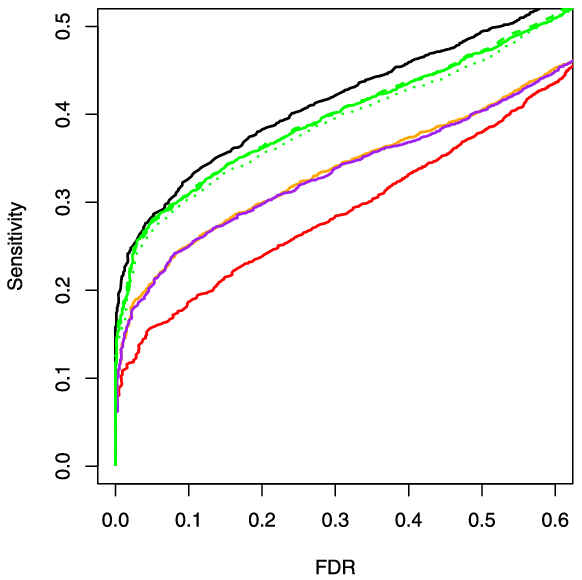}
\\
    (a)   &
    (b)
  \end{tabular}
  \caption{Sensitivity vs. FDR curves in two simulation settings.  \textup{(a)} $f_s=0.1$, $f_g=0.05$, $f_d=0$,
  $n=40$.
    \textup{(b)} $f_s=0.05$, $f_g=0.1$, $f_d=0.1$, $n=80$.}\label{figSensitivityFDR}
\end{figure}

In this section we examine the performance of the proposed BMA approach in
comparison with the single model and the gold standard multi-model
approaches. To evaluate the effects of prior choice on the performance
of the BMA approach, we considered three prior model probability
choices: the proposed empirical prior obtained using the two step
approach (${\mathit{BMA}}_1$), the uniform prior (${\mathit{BMA}}_2$), and the true
proportion of genes for each model ($\mathit{BMA}_3$). The posterior inclusion
probability of $s$ was used as the ranking statistics. The number of
genes identified by each methods at 5\% FDR were compared in
Table~\ref{tablepower}. We can see that the $\mathit{BMA}$ approaches had
greater power in detecting DE genes compared to the $\mathit{SM}$ approaches in
general and the performance came close to that of the $\mathit{MM}$
approach. In fact, in all the simulated settings, the $\mathit{BMA}$
approaches, particularly $\mathit{BMA}_1$ and $\mathit{BMA}_3$, showed sensitivity close
to the $\mathit{MM}$ approach for a given FDR threshold and greater than the
single model approaches. Figure~\ref{figSensitivityFDR} showed the magnitude of
performance difference in two representative settings. The $\mathit{BMA}$
approaches appeared to be relatively insensitive to the choice of prior
model probabilities for gene ranking.
%{\renewcommand{\baselinestretch}{1}
%t2 ###
\begin{table}
\tabcolsep=0pt
    \caption{Power of different methods for identifying genes
      differentially expressed between smokers and nonsmokers at 5\%
      FDR under different simulation settings}
    \label{tablepower}
    \begin{tabular*}{\textwidth}{@{\extracolsep{\fill}}lccccccccc@{}}
      \hline
      $\boldsymbol{f_s}$&$\boldsymbol{f_g}$&$\boldsymbol{f_d}$&$\boldsymbol{{\mathit{SM}}_1}$&$\boldsymbol{{\mathit{SM}}_2}$
      &$\boldsymbol{\mathit{SVA}}$&$\boldsymbol{{\mathit{BMA}}_1}$&$\boldsymbol{{\mathit{BMA}}_2}$&$\boldsymbol{{\mathit{BMA}}_3}$&$\boldsymbol{\mathit{MM}}$\\
      \hline
      \multicolumn{10}{@{}l@{}}{$n=40$}\\
      $0.10$& $0.05$& $0$\hphantom{.05}   &139&\hphantom{0}96&\hphantom{0}83&145&119&149&155\\
      $0.10$& $0.05$& $0.05$&126&\hphantom{0}80&\hphantom{0}72&137&124&137&150\\
      $0.05$& $0.10$& $0$\hphantom{.05}   &\hphantom{0}42&\hphantom{0}31&\hphantom{0}31&\hphantom{0}51&\hphantom{0}46&\hphantom{0}52&\hphantom{0}52\\
      $0.05$& $0.10$& $0.10$&\hphantom{0}46&\hphantom{0}30&\hphantom{0}26&\hphantom{0}56&\hphantom{0}49&\hphantom{0}57&\hphantom{0}58\\
      \multicolumn{10}{@{}l@{}}{$n=80$}\\
      $0.10$& $0.05$& $0$\hphantom{.05}   &286&294&290&346&335&344&356\\
      $0.10$& $0.05$& $0.05$&239&250&248&317&308&318&334\\
      $0.05$& $0.10$& $0$\hphantom{.05}   &\hphantom{0}94&113&110&147&135&146&152\\
      $0.05$& $0.10$& $0.10$&\hphantom{0}82&108&106&145&138&142&147\\
      \hline
    \end{tabular*}
\end{table}
%
%}

%{\renewcommand{\baselinestretch}{1}
%f1 ###
%}

\begin{figure}
  \centering
  \begin{tabular}{cc}

\includegraphics{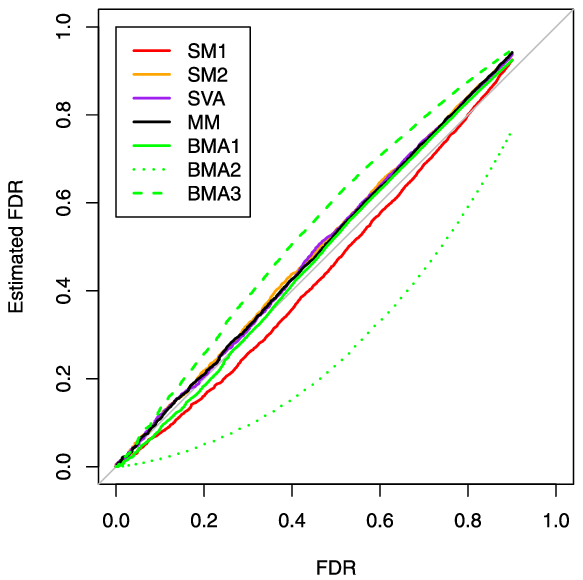}
 &

\includegraphics{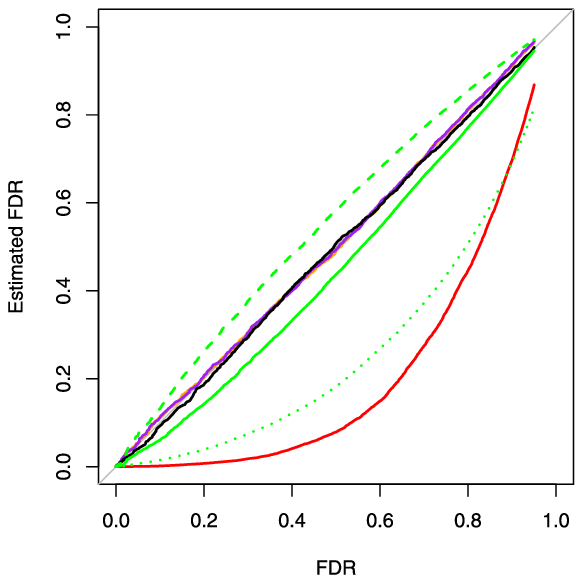}
\\
    (a)   &
    (b)
  \end{tabular}
  \caption{Estimated FDR vs. true FDR in two simulation settings. \textup{(a)} $f_s=0.1$, $f_g=0.05$, $f_d=0$,
  $n=40$.
    \textup{(b)} $f_s=0.05$, $f_g=0.1$, $f_d=0.1$, $n=80$.}
  \label{figfdrest}
\end{figure}

Besides providing proper ranking of the gene, it is often useful to
estimate the FDR of the finding and quantifying the proportion of DE
genes in the transcriptome. Therefore, we also evaluated how well the
FDR could be estimated based on the ranking statistics. For the $p$-value
based approach, FDR and the proportion of DE genes were estimated
using the approach by \citet{Storey2002} and
\citet{Storey2003}. For
the Bayesian model averaging approach, the \textit{pe}FDR was directly
estimated based on the posterior inclusion
probability [\citet{NewtonBiostat2004}], that is,
\[
peFDR_{k}(p)=\sum_j (1-P_{kj})\cdot {\bf 1}_{[P_{kj}\leq p]}\big/\sum_j {\bf
  1}_{[P_{kj}\leq p]},
\]
where $0<p \leq 1$ and $P_{kj}$ is the posterior inclusion
probability of variable $k$ for gene $j$. Figure~\ref{figfdrest}
shows the estimated FDR vs. the true FDR in two representative
settings. We can see that using $p$-values from $\mathit{SM}_1$ in studies with
confounder associated genes, the estimated FDR was smaller than the
true FDR. The magnitude of underestimation increased with the sample
size and the proportion of the confounder associated genes. On the
other hand, the FDR estimated using $p$-values from $\mathit{SM}_2$,
$\mathit{SVA}$ or $\mathit{MM}$ was
very close to the true FDR. The accuracy of the $pe$FDR, as observed
by other researchers, appeared to be sensitive to the prior
choice. $pe$FDR obtained based on $\mathit{BMA}_2$, the Bayesian model
averaging approach with uniform prior can greatly underestimate the
FDR. $pe$FDR obtained based on $\mathit{BMA}_1$ showed improved accuracy in FDR
estimation. The results from our simulation also suggest that
the $pe$FDR based on $\mathit{BMA}_1$ are close to true FDR in all simulated
settings. $\mathit{BMA}_3$ appeared to result in $pe$FDR that slightly
overestimated the FDR. Level of sensitivity of the $\mathit{BMA}_1$
approach to the choice of $c$ and model space misspecification can
be found in \citet{Zhou2011}.
%f2 ###

We also carried out two additional sensitivity analyses
related to the BMA approaches using the simulated microarray data [see
\citet{Zhou2011} for detail]. First, we investigated the
sensitivity of the performance of the empirical BMA
approach to the choice of the cutoff $c$. The results suggest that
the BMA approach with empirical prior is relatively robust in gene
ranking with respect to the value of $c$. Second, we investigated the
performance of the BMA approach to the misspecification of model
space, that is, omission of an important covariate $d$. As expected, there is
a~decrease in ranking performance, but the BMA approach still
outperforms all the single model approaches. It is possible to avoid
the performance loss due to omission of important covariates by
introducing the surrogate variables [\citet{Leek2007PLoS}] into the
models. However, including the surrogate variables in the BMA
approach is not a trivial extension due to model uncertainty, and it
is definitely an interesting future research topic.

%
%s5 ###
\section{Application to the observational micorarray data sets}\label{sec5}
We applied the BMA approach to two smoking related observational
microarray studies. Through the application, we intended
to demonstrate the complex relationship between the gene expression
pattern and sample characteristics and the flexibility of the
BMA approach in capturing and quantifying such
relation in a unified and coherent framework.

%
%s5.1 ###
\subsection{Microarray study of airway epithelium samples}\label{sec5.1}
The first data set (GSE10006) came from a study with a
  total of 87 current and never smokers
   [\citet{Carolan2008Decreased}]. The microarray analyses
were carried out on airway epithelium\vadjust{\goodbreak} samples from these subjects. The data
were preprocessed with the Affymetrix MAS method. After excluding gene
probe sets whose expression measurements were deemed
absent or marginal among all subjects, the remaining data consisted of
expression profiles of 44\mbox{,}085 probe sets of genes from
the Affymetrix HGU133plus2 chip for each
subject. Among these probe sets, 34\mbox{,}614 were annotated
  for probing the expression of 17\mbox{,}690 genes. About half of these genes
  were probed by multiple probes. To eliminate the
  potential dependence issue, average expression measurements were
  obtained for genes with multiple probe sets. We analyzed the
  expression data of the 17\mbox{,}690 genes from 60 healthy subjects. Individuals
with known lung disease were excluded. Besides smoking status,
information on age, gender, race and site of the tissue was available. The
samples were heavily unbalanced, the proportion of smokers
was greater in female participants than in males (86\% vs. 57\%), the
proportion of large airway samples was slightly larger in females than
in males (57\% vs. 46\%), and the proportion of caucasian participants
was larger in females compared to males (43\% vs. 37\%).

\begin{figure}

\includegraphics{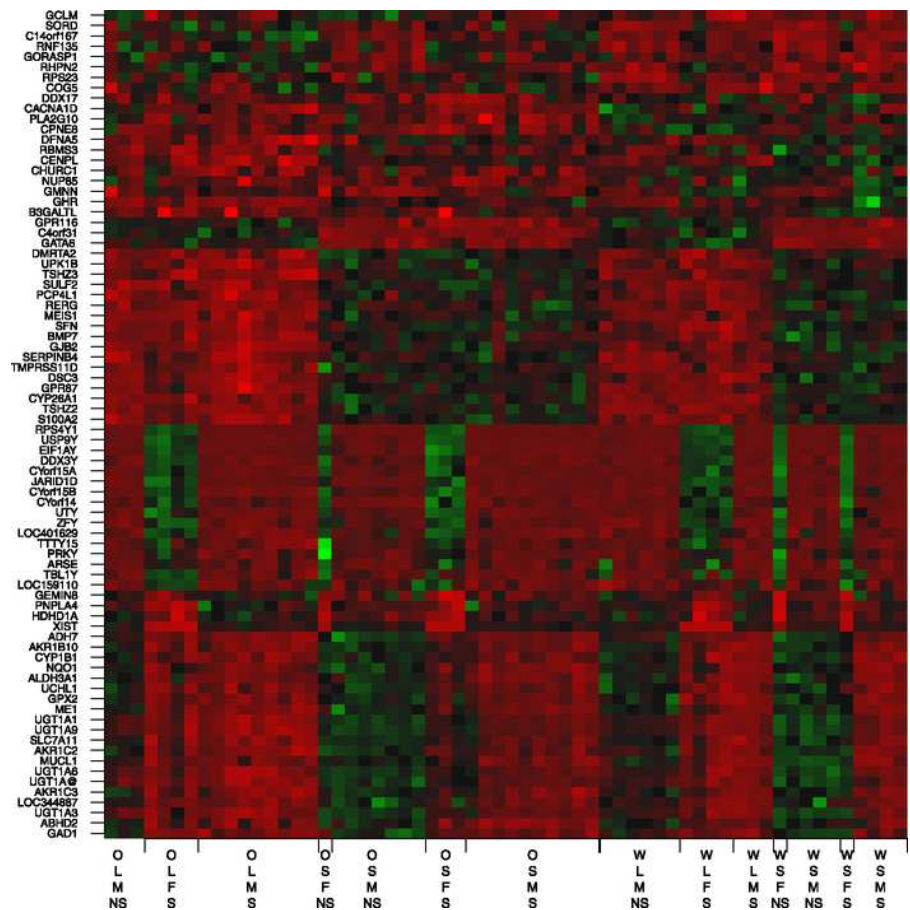}

  \caption{Gene expression intensities for the top 20 genes associated
  with each of the four covariates ($\mathit{smoking}$, $\mathit{gender}$, $\mathit{site}$ and
  $\mathit{race}$) identified by using $\mathit{BMA}_1$. Labels along the x-axis show
  the characteristics of a sample subgroup. From top to bottom, the
  label represents categories of $\mathit{race}$ (Others vs. White; O vs. W),
  $\mathit{site}$ (Large airway vs. Small airway; L vs. S), $\mathit{gender}$ (Male
  vs. Female; M vs. F) and $\mathit{smoking}$ status (Non-Smoker vs. Smoker;
  NS vs. S). For example ``OLMNS'' indicates the subgroup  with the
  following characteristics: Other races (i.e., nonwhite), Large
  airway sample, Male, Non-Smoker.}
  \label{figcrystaltopgeneimage}
\end{figure}

With five covariates, a total of $2^5$ models were
included in the model space. Interaction terms were ignored. The BMA
approach allowed for simultaneous assessment of the association
between the gene expression and each of the sample characteristics,
and straightforward estimation of both the total
proportion of the DE genes in the transcriptome and the proportion of
DE genes associated with each covariate based on Bayes factors. The
application showed a complex picture of the expression pattern in the
epithelium microarray study. A total of 69\% of the genes were estimated to be
differentially expressed. The estimated proportions of DE genes for
association with $\mathit{smoking}$, $\mathit{site}$, $\mathit{gender}$, $\mathit{race}$ and $\mathit{age}$ were
19\%, 34\%, 6\%, 6\% and 4\%, respectively. By
controlling the \textit{pe}FDR at 5\%, we identified a number of DE genes
associated with $\mathit{smoking}$ (928), $\mathit{site}$ (3089), $\mathit{gender}$ (73), $\mathit{race}$
(33) and $\mathit{age}$ (7). The complex expression patterns were illustrated in
Figure~\ref{figcrystaltopgeneimage} where we show the expression
pattern of the top 20 genes associated with $\mathit{smoking}$, $\mathit{gender}$,
$\mathit{site}$ and $\mathit{race}$, respectively.
%{\renewcommand{\baselinestretch}{1}
% Crystal data: top genes expression image
%f3 ###
%}

The results also revealed complex roles of some of these DE genes
which were strongly associated with multiple sample
characteristics. For example, among the top 928 $\mathit{smoking}$ related DE
genes, 343, 18 and 6 of them were also identified as
hits for association with tissue $\mathit{site}$, $\mathit{gender}$ and $\mathit{race}$,
respectively. Additionally, there were 25 genes
identified as hits for association with three or more sample
characteristics, mostly $\mathit{smoking}$, $\mathit{site}$ and $\mathit{gender}$. The BMA
approach allows for assessing jointly genes' association with multiple
sample characteristics. For example, the joint posterior inclusion
probability of $\mathit{smoking}$, $\mathit{site}$ and $\mathit{gender}$ can be obtained by
summing  over the posterior probabilities of models containing all
three covariates. \textit{pe}FDR can then be derived similarly using this
posterior inclusion probability. The analysis identified
6 genes, NRARP, TMEM178, UGT1A@, UGT1A1, UGT1A3 and
UGT1A6, as hits for joint association with the three characteristics
at 5\% $pe$FDR. The existence of such genes suggested a connection
between tobacco smoking and the functions of these genes which were
partly revealed through their association with the phenotype of the
subjects from whom samples were obtained. Results from such analysis
offer additional important information that is useful for generating
new hypotheses and insights into the effects of tobacco smoke on the
transcriptome.

As discussed in the previous sections, given the existence of genes
associated with various sample characteristics, single model
approaches were subjected to the effects of increased bias or reduced
power in unbalanced study design. For the epithelium microarray data,
we saw large differences in gene rankings derived from the
BMA approach and the three single model approaches.
Among the top 1000 smoking related DE genes identified by
each method, the agreement was merely 19.7\% among all four
methods. Specifically, the $\mathit{SVA}$ approach produced gene lists that
were vastly different from the gene lists produced by the other
approaches, where more than half of the top 1000 genes had ranks
beyond 1000 by the other three methods [see the Venn diagram in
\citet{Zhou2011}]. Careful examination of the
gene lists produced
by the $\mathit{SVA}$ approach suggested possible effects of overfitting as
the $\mathit{SVA}$ approach adjusted for a total of 10 surrogate
variables for each gene. The agreement was about 56\%
for the $\mathit{SM}_1$, $\mathit{SM}_2$ and $\mathit{BMA}_1$ approaches, that is, 56\% were ranked
within the top 1000 by all three methods. The agreement between
$\mathit{BMA}_1$ and each of the single model approaches ($\mathit{SM}_1$, $\mathit{SM}_2$ and
$\mathit{SVA}$) was 85\%, 64\% and 35\%, respectively. These differences
were driven by the genes whose expression patterns were not properly
captured by the single model. The higher agreement between results
from $\mathit{BMA}_1$ and $\mathit{SM}_1$ reflects the fact that a large
proportion of the $\mathit{smoking}$ related DE genes are associated with
$\mathit{smoking}$ only.

%s5.2 ###
\subsection{Microarray study of oral mucosa samples}\label{sec5.2}
The second data set came from a study with a total of 79 age and gender
matched healthy smokers and never smokers
[\citet{BoyleGumusDannenberg2010}]. The microarray analyses
were carried out on oral mucosa samples obtained from these subjects
through buccal biopsies. The preprocessed microarray data consisted of
24\mbox{,}103 probe sets of genes from the Affymetrix HGU133plus2 chip
for each subject. Among these probe sets, 22\mbox{,}004 were annotated
for probing the expression of 12\mbox{,}911 genes. About 43\% of these genes
were probed by multiple probe sets. To eliminate the potential
dependence issue, average expression measurements again were
obtained for these genes. The analysis was carried for the
expression data of the 12\mbox{,}911 genes. For subjects recruited for this
study, information regarding age, gender and smoking status was available.

The study samples were balanced in terms of gender between smokers and
nonsmokers. Therefore, single model approaches with or without
adjustment for gender would provide similar results. However, one
interesting biological question was whether there were genes affected
by smoking differently between the males and females. In this context,
direct application of the single model approach could lead to
confusing results. For example, at 5\% estimated FDR, the single model
without adjustment for the interaction term resulted in
944 hits for association with $\mathit{smoking}$, while the
model adjusted for both $\mathit{gender}$ and $\mathit{gender} \times \mathit{smoking}$
interaction led to the identification of only 1 gene
as hits for association with $\mathit{smoking}$ and no genes were identified as
hits for $\mathit{smoking} \times \mathit{gender}$ interaction. Such large
difference in DE gene assessment between different models is difficult
to reconcile and interpret under the single model framework. Yet, such
difference can be expected if there are genes associated with the
interaction because the two variables, $\mathit{smoking}$ and $\mathit{smoking} \times
\mathit{gender}$ interaction, are correlated. Joint testing of the effects of
$\mathit{smoking}$ and $\mathit{smoking} \times \mathit{gender}$ interaction led to the
identification of 311 DE genes with the likelihood ratio
test. However, this method can not quantify the relative contribution
from the two variables. We therefore applied the BMA approach to these
data to illustrate the flexibility and usefulness of this approach to
handle possible interaction effects.

In this application, the model space consists of a total of 16 models
including the null model, three models with $\mathit{smoking}$ and/or $\mathit{gender}$
as main effects only and 12 models for different patterns that could
arise from interaction between $\mathit{smoking}$ and $\mathit{gender}$. For the oral
mucosa data, our analysis estimated that about 22.5\%
of the genes are differentially expressed, in which about
12.3\%, 1.5\% and 8.6\% were associated with
$\mathit{smoking}$, $\mathit{gender}$ and $\mathit{smoking} \times \mathit{gender}$
interaction, respectively. Controlling the \textit{pe}FDR at 5\%, our
approach identified a total of 414 genes as hits
associated with smoking through either the main effect, the
interaction effect or both. Specifically, 222 of
these genes were associated with $\mathit{smoking}$ primarily through the main effect,
2 were associated with $\mathit{smoking}$ primarily through the interaction
effect, while for the rest of these genes various degrees of
association were contributed from the interaction term.

%{\renewcommand{\baselinestretch}{1}
%t3 ###
\begin{sidewaystable}
\tabcolsep=0pt
\tablewidth=\textwidth
  \caption{Posterior inclusion probabilities of a
  single covariate, $s$ (for $\mathit{smoking}$), $g$ (for $\mathit{gender}$), or $s
  \times g$ interaction, and a composite of covariates, $s$ and/or $s
  \times g$ interaction (denoted as $s|s \times g$), obtained under
  $\mathit{BMA}_1$, for a list of DE genes associated with $s$ primarily
  through $s \times g$ interaction. Also shown are the ranks of these
  genes based on the strength of association with the covariate$/$s
  under different methods ($\mathcal{R}^{method}_{\mathit{covariate}/s}$)}
  \label{tabDannenberggenelist}
  \begin{tabular*}{\textwidth}{@{\extracolsep{\fill}}lccccce{4.0}e{4.0}e{4.0}e{4.0}e{2.0}e{4.0}e{3.0}@{}}
    \hline
    \textbf{GSymbol} &\textbf{Cytoband}&$\boldsymbol{P_{s}}$ & $\boldsymbol{P_{g}}$ &
    $\boldsymbol{P_{s \times g}}$ &$\boldsymbol{P_{s|s \times g}}$ &\multicolumn{1}{c}{$\boldsymbol{\mathcal{R}^{\mathit{SM}_1}_{s}}$}&
    \multicolumn{1}{c}{$\boldsymbol{\mathcal{R}^{\mathit{SM}_2}_{s}}$} &\multicolumn{1}{c}{$\boldsymbol{\mathcal{R}^{\mathit{BMA}_1}_{s}}$} &
    \multicolumn{1}{c}{$\boldsymbol{\mathcal{R}^{\mathit{SM}_2}_{s \times g}}$}  &
    \multicolumn{1}{c}{$\boldsymbol{\mathcal{R}^{\mathit{BMA}_1}_{s \times g}}$} &
    \multicolumn{1}{c}{$\boldsymbol{\mathcal{R}^{\mathit{SM}_2}_{s|s \times g}}$}&
    \multicolumn{1}{c@{}}{$\boldsymbol{\mathcal{R}^{\mathit{BMA}_1}_{s|s \times g}}$}\\
    \hline
      CEACAM7& 19q13.2&0.042&0.003&0.969&0.998& 205& 3366& 6703&
       96& 1& 66& 39\\
      CD177  & 19q13.2&0.035&0.006&0.933&0.962&1156& 6912& 8041&
      327& 2& 523&191\\
      MARK1  &    1q41&0.061&0.004&0.928&0.985& 485& 9489& 4967&
      6& 3& 83&122\\
      GTF2A2 & 15q22.2&0.055&0.005&0.904&0.953& 997& 6777& 5367&
      260& 4& 425&214 \\
      PLA2G2A&    1p35&0.092&0.008&0.878&0.963& 643& 3677& 3636&
      805& 5& 375&189\\
      AKR1B10&    7q33&0.062&0.008&0.875&0.931&1128& 5915& 4929&
      608& 6& 618&278\\
      THYN1  &   11q25&0.020&0.033&0.869&0.885&3123&10\mbox{,}582&12\mbox{,}884&
      522&7&1583&384\\
      BMS1   &10q11.21&0.117&0.008&0.861&0.970& 502& 3142& 3017&
      774& 8& 278&169\\
      CLIC2  &    Xq28&0.079&0.029&0.858&0.934& 910& 3395& 4095&
      1929& 9& 686&265\\
      PRDX5  &   11q13&0.059&0.061&0.854&0.908&1266& 3128& 5065&
      3370&10&1004&331\\
      \hline
    \end{tabular*}
\end{sidewaystable}
%}

By comparing the $\mathit{smoking}$ related DE genes identified by the single
model approaches and the BMA approach, we noted
that the difference was mainly from genes that were over/under
expressed in only one subgroup of the subjects, female
smokers. Neither the model with $\mathit{smoking}$ status as the only covariate
nor the full model adjusted for both the $\mathit{gender}$ and the $ \mathit{smoking}
\times \mathit{gender}$ interaction were able to adequately capture the
strength of association for this group of genes and properly rank them
due to either increased bias or decreased
power. Table~\ref{tabDannenberggenelist} showed the
posterior inclusion probabilities and ranks based on different
approaches for a few of these genes. A large difference in the
rankings by different methods can be seen.

%
%s6 ###
\section{Discussion}\label{sec6}
In the past decade, microarray technology has greatly increased our
ability to simultaneously interrogate the expression of tens of
thousands of genes. Use of this technology has contributed to an
improved understanding of the molecular basis of various diseases. As
one of the primary tools for such studies, methods for finding DE genes
have also been refined over time. Various approaches have been
proposed to deal with multiple issues in microarray data. Yet, from
the modeling perspective, many approaches have ignored sample
heterogeneity, its impact on the analysis results, and
the great opportunity it presents. Since
\citet{Potter2003Epidemiology} discussed the need
for controlling bias and confounding in observational microarray
studies, it has been increasingly recognized that the lack of control
for sample heterogeneity could be a barrier to the reproducibility of
the study findings. In two
editorials [\citet{Webb2007Microarrays}; \citet{Troester2009Microarrays}],
improved data analysis methods and better study design
have been considered crucial for advancing the field of cancer
epidemiology with microarray technology. In particular,
\citet{Troester2009Microarrays} discussed the potential of model
selection strategies in the process. Nevertheless, there remain
obstacles to fully appreciate the effect of complex sample
characteristics on DE gene detection and the value of improving upon
current approaches.

In this paper, we proposed a novel concept for high throughput data
analysis involving a heterogeneous sample, that is, a multi-model handling
is intrinsically needed. We presented the theoretical framework that
explains why basing inferences on a single model could be problematic
in observational microarray studies. The problem arises from the
inadequacy of using a single model to describe the complex expression
pattern of genes among a~heterogeneous sample, which can result in
increased number of false discoveries due to bias when a simple model
is used or increased random error due to reduced efficiency when a
complex model is used. Such effects of model misspecification are hard
to avoid because of the existence of genes being affected by
different sets of sample characteristics and/or their
interactions. We showed through simulation that the single model
approaches have inferior performance in DE gene finding in comparison
with a multi-model approach should we know the right model for the
right set of genes. The magnitude of effects on false discovery
depends on the study design, specific biological system and the
mechanism underlying expression variation.

We proposed to use the BMA approach to improve our ability to identify DE
genes. This approach utilizes the Zellner--Siow
prior for model parameters. The consistency property of this prior is
important, as it allows for obtaining a consistent estimate of the
distribution of the genes in the model space using Bayes
factors. Another choice could be the hyper-$g/n$ prior proposed in
\citet{LiangJASA2008}. We proposed to use an iterative procedure to
obtain the prior model probabilities so that the estimated distribution
of the genes among the model space based on posterior model
probabilities matches the estimate based on the Bayes
factors. These prior choices allow the efficient
computation of the Bayes factors and the posterior inclusion
probabilities that does not rely on a MCMC simulation. Our
simulation study demonstrated that this approach performed almost as
well as the gold standard multi-model approach with true models and
better than the single model approaches in gene ranking. The ranking
performance was relatively insensitive to a wide range of choice for
prior model probabilities. However, accuracy of the FDR directly
estimated from the posterior model/inclusion probabilities was
sensitive to the prior choice. Our simulation study showed that the
proposed empirical prior model probability allowed for reasonably good
calibration of posterior model/inclusion probabilities for
multiplicity and the estimated FDR was close to the true FDR in settings with
moderate to large sample size. In the rare case of a small study with
a heterogeneous sample, care needs to be taken when using the
empirical prior because the small sample size property of the
Zellner--Siow prior is less certain. Nevertheless, it should be
pointed out that multiplicity control in the Bayesian modeling
framework remains a challenging and active research area. Further
studies on the theoretical aspects of the prior choice for
multiplicity control
across the multiple genes and multiple models are needed. The current
BMA approach is developed under the M-complete assumption, that is, the
model space contains the true model. Should unknown confounders exist,
it is possible to capture the latent confounding factors by
introducing the surrogate variables [\citet{Leek2007PLoS}]. We note,
however, it would be unwise to directly incorporate the surrogate
variables, currently constructed based on residuals derived from a
single model fit of the data, into the proposed BMA
approaches. Our work relies on the assumption of
  linear regression models with normal errors, which may be violated
  in practice. This calls for new approaches that are
  robust to the normality assumption, which is likely to be
  particularly useful for studies with small sample sizes. For the
  analysis of conventional data with one outcome variable, robust
  Bayesian model selection/averaging approaches have been suggested,
  for example, the approach by \citet{Gottardo2009}. Extending such
  ideas to the observational microarray studies represents an
  interesting future direction.

Finally, through the application of the BMA approach to an
observational mircoarray study with unbalanced study design and one
with balanced study design, we showed that complex expression
patterns did exist when study samples were heterogeneous. Previous research
has demonstrated the complexities of underlying biological mechanisms
for gene expression variation. Genes affected by several common
factors, such as age [\citet{Tan2008Differential}],
gender [\citet{Delongchamp2005Genome}; \citet{Yang2006Gender}; \citet{Tan2008Differential}],
smoking [\citet{Spira2004Effects}] and drinking
alcohol [\citet{Lewohl2001Application}], have been found in different
tissue samples. Our study showed that such complexity interfered
with the DE gene detection. Notably, the BMA approach was able to
avoid missing important genes whose expression patterns were not
adequately captured by a single model approach. As an added value, the
BMA approach is found to be a flexible tool that allows for more
comprehensive characterization of the association between gene
expression and the characteristics of the subjects from whom the
samples were obtained. All these can be done within a
unified and coherent framework.

%%\begin{supplement}[id=supp]
%%  % \sname{Supplement}
%%  % \stitle{A Bayesian Model Averaging Approach for Observational
%%  %  Gene Expression Studies}
%%  \slink[doi]{???}
%%  \slink[url]{http://lib.stat.cmu.edu/aoas/???/???}
%%  \sdatatype{.pdf}
%%  \sdescription{Detailed description of the simulation setup and simulation
%%    procedure and additional results from the simulation study and
%%    application to the airway epithelium microarray study are provided.}
%%\end{supplement}
%
%%

\section*{Acknowledgments}
The authors thank Doctors Jaya Satagopan and Li-Xuan Qin at the Memorial
Sloan-Kettering Cancer Center for helpful
discussions. The authors are grateful to the Editor,
  the Associate Editor and four anonymous referees whose comments and
  suggestions greatly improved this article. {\textit{Conflict of Interest:} None declared.}

\begin{supplement}%[id=suppA]
\stitle{Supplement to ``A Bayesian model averaging approach for
  observational gene expression studies''}
\slink[doi]{10.1214/11-AOAS526SUPP}  %[doi,text={...}] - jei reikia %suskaldyti doi
\slink[url]{http://lib.stat.cmu.edu/aoas/526/supplement.pdf}
\sdatatype{.pdf}
\sdescription{Detailed description of the simulation setup and simulation
   procedure and additional results from the simulation study and
   application to the airway epithelium microarray study are provided.}
\end{supplement}

% imsref loaded by smiklovaite, 2012-01-23 07:42:12
% imsref loaded by smiklovaite, 2012-01-23 07:59:21

\printaddresses

\end{document}